\documentclass[traditabstract]{aa}
\usepackage{txfonts}
\usepackage{graphicx}
\usepackage{natbib}
\bibpunct{(}{)}{;}{a}{}{,} 
\usepackage{epsfig}

\newcommand{\teff}{\mbox{$T_{\rm eff}$}}
\newcommand{\logg}{\mbox{$\log g$}}
\newcommand{\vsini}{\mbox{$v \sin I$ }}
\newcommand{\mictrb}{\mbox{$\xi_{\rm t}$}}
\newcommand{\mactrb}{\mbox{$v_{\rm mac}$}}

\newcommand{\halpha}{\mbox{$H_\alpha$}}

\begin{document}
\title{WASP-23b: a transiting hot Jupiter around a K dwarf\thanks{using WASP-South photometric observations confirmed with LCOGT Faulkes South Telescope, the 60\,cm TRAPPIST telescope, the \textit{CORALIE} spectrograph and the camera from the Swiss 1.2\,m \textit{Euler} Telescope placed at La Silla, Chile, as well as with the HARPS spectrograph, mounted on the ESO 3.6m, also at La Silla, under proposal 084.C-0185. The data is publicly available at the \textit{CDS} Strasbourg and on demand to the main author.} 
\\ {\large  and its Rossiter-McLaughlin effect}}


\author{Amaury H.M.J. Triaud\inst{1}
\and Didier Queloz\inst{1}
\and Coel Hellier\inst{2}
\and Micha\"el Gillon\inst{3} 
\and Barry Smalley\inst{2}
\and Leslie Hebb\inst{4}
\and Andrew Collier Cameron\inst{5}
\and David Anderson\inst{2}
\and Isabelle Boisse\inst{6,7}
\and Guillaume H\'ebrard\inst{7,8}
\and Emmanuel Jehin\inst{3}
\and Tim Lister\inst{9}
\and Christophe Lovis\inst{1}
\and Pierre F.L. Maxted\inst{2}
\and Francesco Pepe\inst{1}
\and Don Pollacco\inst{10}
\and Damien S\'egransan\inst{1}
\and Elaine Simpson\inst{10}
\and St\'ephane Udry \inst{1}
\and Richard West\inst{11}
}

\offprints{Amaury.Triaud@unige.ch}

\institute{Observatoire Astronomique de l'Universit\'e de Gen\`eve, Chemin des Maillettes 51, CH-1290 Sauverny, Switzerland
\and Astrophysics Group, Keele University, Staffordshire, ST55BG, UK
\and Institut d'Astrophysique et de G\'eophysique, Universit\'e de Li\`ege, All\'ee du 6 Ao\^ut, 17, Bat. B5C, Li\`ege 1, Belgium
\and Department of Physics and Astronomy, Vanderbilt University, Nashville, TN37235, USA
\and SUPA, School of Physics \& Astronomy, University of St Andrews, North Haugh, KY16 9SS, St Andrews, Fife, Scotland, UK
\and Centro de Astrof\'isica, Universidade do Porto, Rua das Estrelas, 4150-762 Porto, Portugal
\and Institut d'Astrophysique de Paris, CNRS (UMR 7095), Universit\'e Pierre \& Marie Curie, 98bis bd. Arago, 75014 Paris, France
\and Observatoire de Haute-Provence, CNRS/OAMP, 04870 St Michel l'Observatoire, France
\and Las Cumbres Observatory, 6740 Cortona Dr. Suite 102, Santa Barbara, CA 93117, USA
\and Astrophysics Research Centre, School of Mathematics \& Physics, QueenÕs University, University Road, Belfast, BT71NN, UK
\and Department of Physics and Astronomy, Universityof Leicester, Leicester, LE17RH, UK
}

\date{Received date / accepted date}
\authorrunning{Triaud et al.}
\titlerunning{WASP-23b, a transiting planet}

\abstract{We report the discovery of a new transiting planet in the Southern Hemisphere. It has been found by the WASP-south transit survey and confirmed photometrically and spectroscopically by the 1.2m Swiss \textit{Euler} telescope, LCOGT 2m Faulkes South Telescope, the 60\,cm TRAPPIST telescope and the ESO 3.6m telescope. The orbital period of the planet is 2.94 days. We find it is a gas giant with a mass of $0.88\pm 0.10\,M_\mathrm{J}$ and a radius estimated at $0.96 \pm 0.05\,R_\mathrm{J}$. We have also obtained spectra during transit with the HARPS spectrograph and detect the Rossiter-McLaughlin effect despite its small amplitude. Because of the low signal to noise of the effect and of a small impact parameter we cannot place a constraint on the projected spin-orbit angle. We find two conflicting values for the stellar rotation. Our determination, via spectral line broadening gives $v\,\sin\,I = 2.2 \pm 0.3$ km\,s$^{-1}$, while another method, based on the activity level using the index $\log\,R'_{HK}$,  gives an equatorial rotation velocity of only $v  = 1.35 \pm 0.20$ km\,s$^{-1}$. Using these as priors in our analysis, the planet could either be misaligned or aligned. This should send strong warnings regarding the use of such priors. There is no evidence for eccentricity nor of any radial velocity drift with time.
 }

\maketitle

\section{Introduction}

Finding planets by their transit has proved to be very successful with detections past the hundred mark. After the discovery that HD\,209458b was transiting \citep{Charbonneau:2000p5431}, a plethora of ground based small aperture wide-angle photometric surveys were put in place to find such bodies such as WASP \citep{Pollacco:2006p1500}, the HAT network \citep{Bakos:2004p1723}, XO \citep{McCullough:2005p1734}, TrES \citep{ODonovan:2006p1736} or the OGLE search \citep{Udalski:1997p1770, Snellen:2007p1778}. WASP is the only survey currently operating in both hemispheres. About 20\,\% of discovered extrasolar planets are currently known to transit their host stars, the vast majority of which are the so called \textit{hot Jupiters}, planets similar in mass to Jupiter but on orbits with periods $< 5$ days. 

Transiting planets bring a treasure trove of observables allowing the study of a special class of planets,
a class which is absent from our Solar System. 
A transiting system, observed with photometry and radial velocities, allows a measure of  the planet's mass ratio with the star, gives a measurement of the stellar density and ratio of radii. Through observations at the time of occultation, it is possible to obtain indications about the temperature of the planet. Careful analysis during transit and occultation can give clues to its atmospherical composition. 

Finally, another observable has been under intense scrutiny recently: the spin-orbit angle. Indeed, as the planet transits, it will cover a part of the approaching or receding hemisphere of the star, therefore red-shifting or blue-shifting the spectrum. This appears as a radial velocity anomaly in the main reflex Doppler motion curve. It is called the Rossiter-McLaughlin effect  \citep{Holt:1893,Rossiter:1924p869,McLaughlin:1924p872} and was first measured for a planet by \citet{Queloz:2000p247} and modelled by \citet{Ohta:2005p631}, \citet{Gimenez:2006p31} and \citet{Hirano:2010p5304}. Recently it was found that hot Jupiters can be found on a vast range of orbital planes with respect to the stellar rotation, some even on retrograde orbits \citep{Hebrard:2008p226,Winn:2009p3712,Narita:2009p5188,Anderson:2010p5177,Queloz:2010p7376}. The study of this angle's distribution is being used to distinguish the processes through which hot Jupiters have arrived at their current orbits \citep{Triaud:2010p8039, Winn:2010p7311, Morton:2010p8042}.

All this gathered data helps developments in theoretical physics in regimes beforehand out of reach: intense heat transfer between hot and cold hemispheres \citep{Guillot:2002p7421} and on supersonic winds \citep{DobbsDixon:2010p7385} to name only two. The few detections of multi-planet systems in which at least one component is transiting can also informs us about the interior structure \citep{Batygin:2009p7407}. Obviously the study of these special exoplanets is also shedding light on how planets form as well as on the evolution of their orbits with time. The hot Jupiters are thought to have experienced a migration to the star after their formation beyond the ice-line, be it through some angular moment exchange with the primordial protoplanetary disc \citep{Lin:1996p5847}, or via dynamical interactions and subsequent tidal friction \citep{Fabrycky:2007p3141,Nagasawa:2008p2997,Malmberg:2010p7427}, an explanation now preferred to the previous one. The history of that post formation evolution might hold a key to the understanding of the various processes that planetary systems are likely to experience and thus, shed a light on the events around the origin of our own Solar System. 

In that light, we announce the discovery of a new transiting gas giant by the WASP consortium, in close proximity with its host star, another brick in the understanding of these objects.

\section{Observations}\label{sec:obs}

\begin{figure}
\centering                     
\includegraphics[width=9cm]{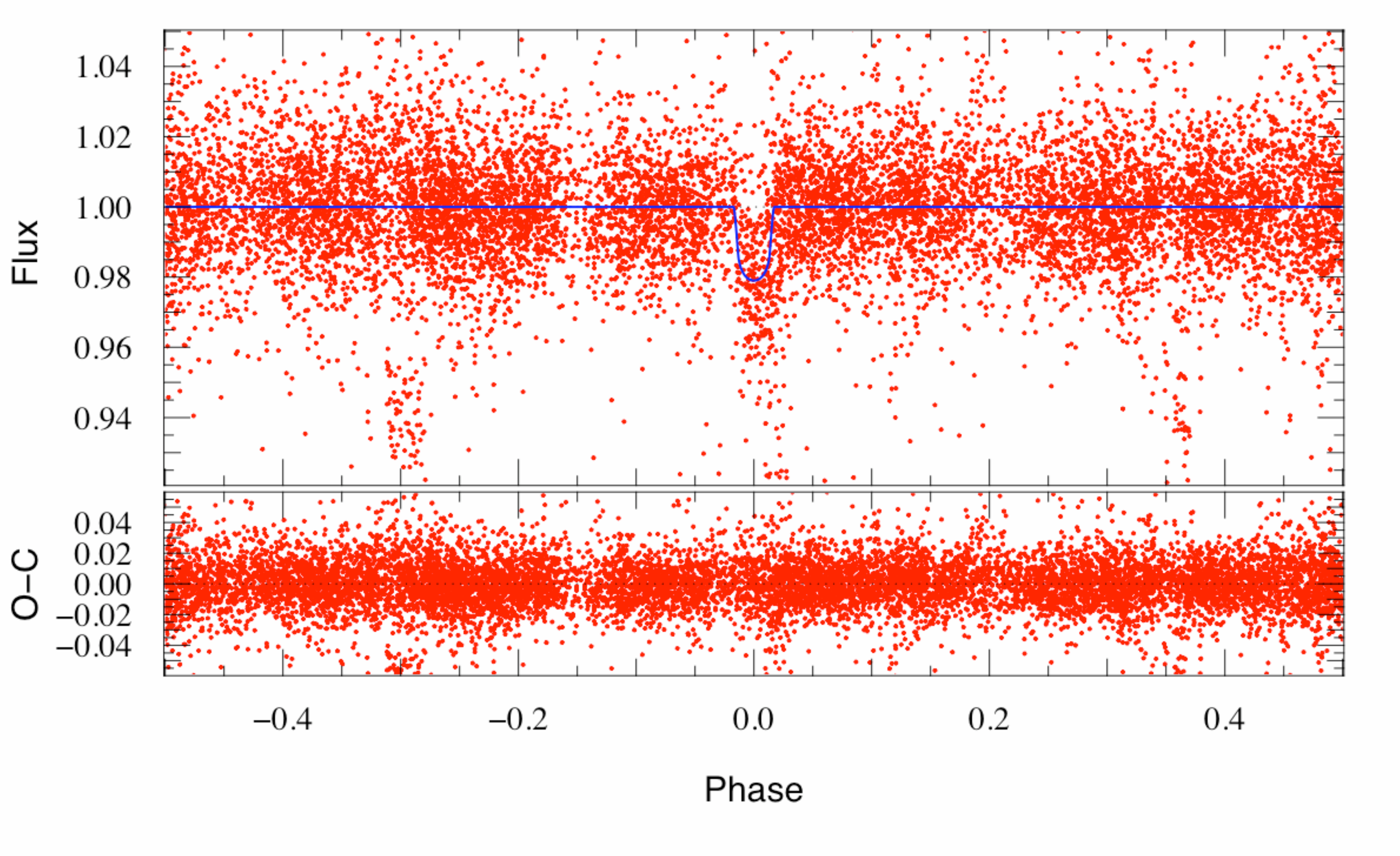}
\caption{Phased WASP-South photometry, of two seasons, and residuals. \textit{R} band model superimposed.}\label{fig:SW}

\end{figure}

The object, WASP-23, (1SWASP J064430.59-424542.5) is a K1V star with V=12.68 
and was observed through two seasons of the WASP-South survey, located in Sutherland, South Africa, in a single camera field from  2006 October 13 to 2007 March 11 and from 2007 October 11 to 2008 March 11 representing 10\,846 photometric measurements. The WASP-South instrument, part of the WASP survey is amply described in \citet{Pollacco:2006p1500}. The \textit{Hunter} algorithm \citep{CollierCameron:2007p704} searched the data and found 11 partial transits with a period of 2.94 days and a depth of 1.7\,\% in both seasons (Fig.~\ref{fig:SW}). It was classed for spectroscopic follow-up.  No rotational variability could be found in the photometric data, indicating slow rotation and few stellar spots.  

\begin{table}[h]
\caption{Stellar parameters of WASP-23 from spectroscopic analysis.}\label{wasp23-params}

\begin{tabular}{llcll}
\hline
\hline
\teff      & 5150 $\pm$ 100 K              &&{[Fe/H]}   &$-$0.05 $\pm$ 0.13 \\
\logg      & 4.4 $\pm$ 0.2                    &&{[Mg/H]}   &  +0.15 $\pm$ 0.15 \\
\mictrb    & 0.8 $\pm$ 0.2 km\,s$^{-1}$          &&{[Si/H]}   &  +0.03 $\pm$ 0.08 \\
\mactrb  & 0.8 $\pm$ 0.3 km\,s$^{-1}$           &&{[Ca/H]}   &  +0.17 $\pm$ 0.16 \\
\vsini     & 2.2 $\pm$ 0.3 km\,s$^{-1}$            &&{[Sc/H]}   &  +0.03 $\pm$ 0.12 \\
                            &                                 &&{[Ti/H]}   &  +0.18 $\pm$ 0.13 \\
B-V & $0.88 \pm 0.05$                       &&{[V/H]}    &  +0.34 $\pm$ 0.13 \\
log $R'_\mathrm{HK}$ &$-4.68\pm0.07$&&{[Cr/H]}   &  +0.04 $\pm$ 0.10 \\
$S_\mathrm{MW}$&$0.32\pm0.04$&&{[Mn/H]}   &  +0.05 $\pm$ 0.15 \\
                            &                                 &&{[Co/H]}   &  +0.11 $\pm$ 0.15 \\
  log A(Li)  &$<$1.0                             &&{[Ni/H]}   &$-$0.03 $\pm$ 0.12 \\
\hline
\\
\end{tabular}
\end{table}

\begin{table}[h]
\caption{Limb darkening coefficients used (quadratic law)}\label{tab:limb}

\begin{tabular}{lllllll}
\hline
\hline
Band & $u_a$ & $u_b$&&Band & $u_a$ & $u_b$\\
\hline 
$V_\mathrm{HARPS}$      & 0.576      & 0.191    & &$z$    & 0.284 & 0.289\\
$R$      & 0.450         & 0.260                          &&$I+z$  & 0.325   &  0.275   \\

\\
\end{tabular}
\end{table}

\begin{figure}
\centering                     
\includegraphics[width=9cm]{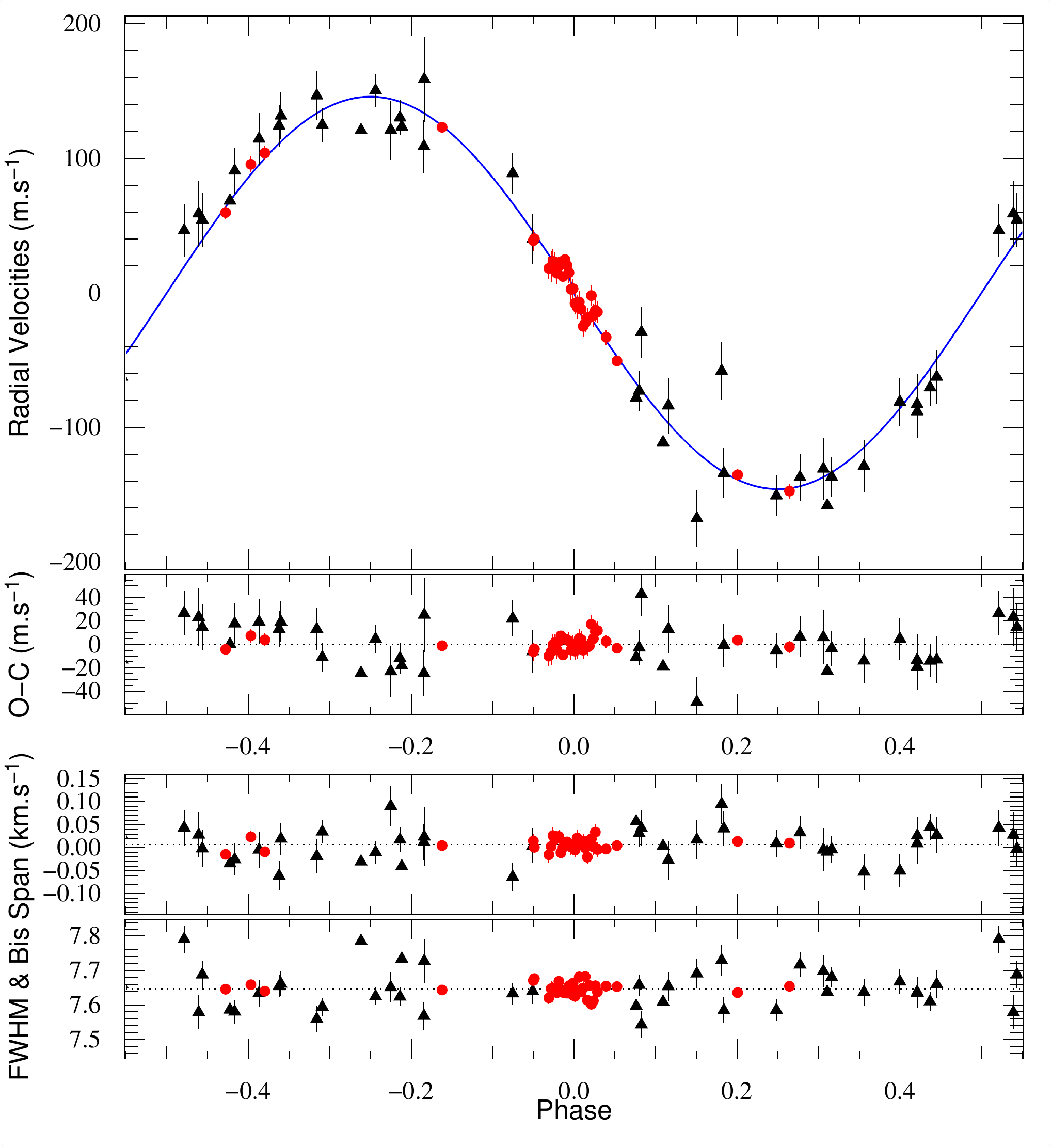}
\caption{Black triangles: CORALIE data; red discs: HARPS data. \textit{top}: radial velocities with model superimposed, and residuals (both in m\,s$^{-1}$), as a function of orbital phase. Added are the $1\,\sigma$ error bars. 
\textit{bottom}: Phased bisector span and FWHM (both in km\,s$^{-1}$). The HARPS data has been translated to have its mean correspond to the CORALIE data.}\label{fig:RV}

\end{figure}

The 1.2\,m \textit{Euler} Swiss telescope, in La Silla, Chile, established the planetary nature of the object by observing the presence of a Doppler variation of semi-amplitude $145\,\mathrm{m\,s}^{-1}$ with the same period and epoch as the WASP-South photometry. Observations started on 2008 August 31 and were pursued until 2010 April 08 totalling 38 radial velocity measurements (Fig.~\ref{fig:RV}), each of 30 minute exposure.

A photometric timeseries was acquired with the camera mounted on the \textit{Euler} telescope, in the $z$-band on 2008 December 13. We gathered 254 measurements and confirmed the reality of the photometric signal discovered by WASP-South (Fig.~\ref{fig:euler}). In addition we gathered another 215 measurement in the $z$ band during transit with the Faulkes South Telescope on 2009 September 27.
We observed a third transit on 2010 February 7, with \textit{Euler}, collecting 193 datapoints in the $R$-band filter (Fig.~\ref{fig:euler}). Two further transits were observed during December 2010 using the newly built 60\,cm TRAPPIST robotic telescope \citep{Gillon:2011p8399}, also located in La Silla, was finally added to this analysis (Fig. \ref{fig:trappist}).

Under ESO proposal 084.C-0185 we observed with the spectrograph HARPS, mounted on the ESO 3.6m telescope, at La Silla, Chile, obtaining 35 spectra between 2009 December 18 and  2010 February 9. 28 spectra at a mean cadence of roughly 600s were acquired on the first night,  14 of which are positioned while the planet transits (Fig.~\ref{fig:RV} \& \ref{fig:euler}). The others were observed a few months later because of scheduling contraints and have exposure times of 1200s.


\section{Data Analysis}\label{sec:analysis}


\subsection{the \textit{Euler} $z$-band transit}\label{subsec:ztransit}

The transit was observed in  the $z$-band, on 2008 December 12, from 2h15 to 7h35 UTC using  the \textit{Euler} camera. The $z$-band filter was used to minimise the impact of stellar limb-darkening on the deduced system parameters. The images were $2 \times 2$ binned to improve the duty cycle of the observations, resulting in a pixel scale of 0.7 arcsec. 254 exposures were acquired during the run, with an exposure time ranging from 45s to 60s.  Two outliers were removed for the analysis.
To keep a good spatial sampling while minimising the impact of interpixel sensitivity inhomogeneity and  seeing variations, the telescope was heavily defocused: the mean profile width was $4.8 \pm 0.2$ arcsec. 
Airmass decreased from 1.43 to 1.03 then raised to 1.09. 

\begin{figure}
\centering                     
\includegraphics[width=9cm]{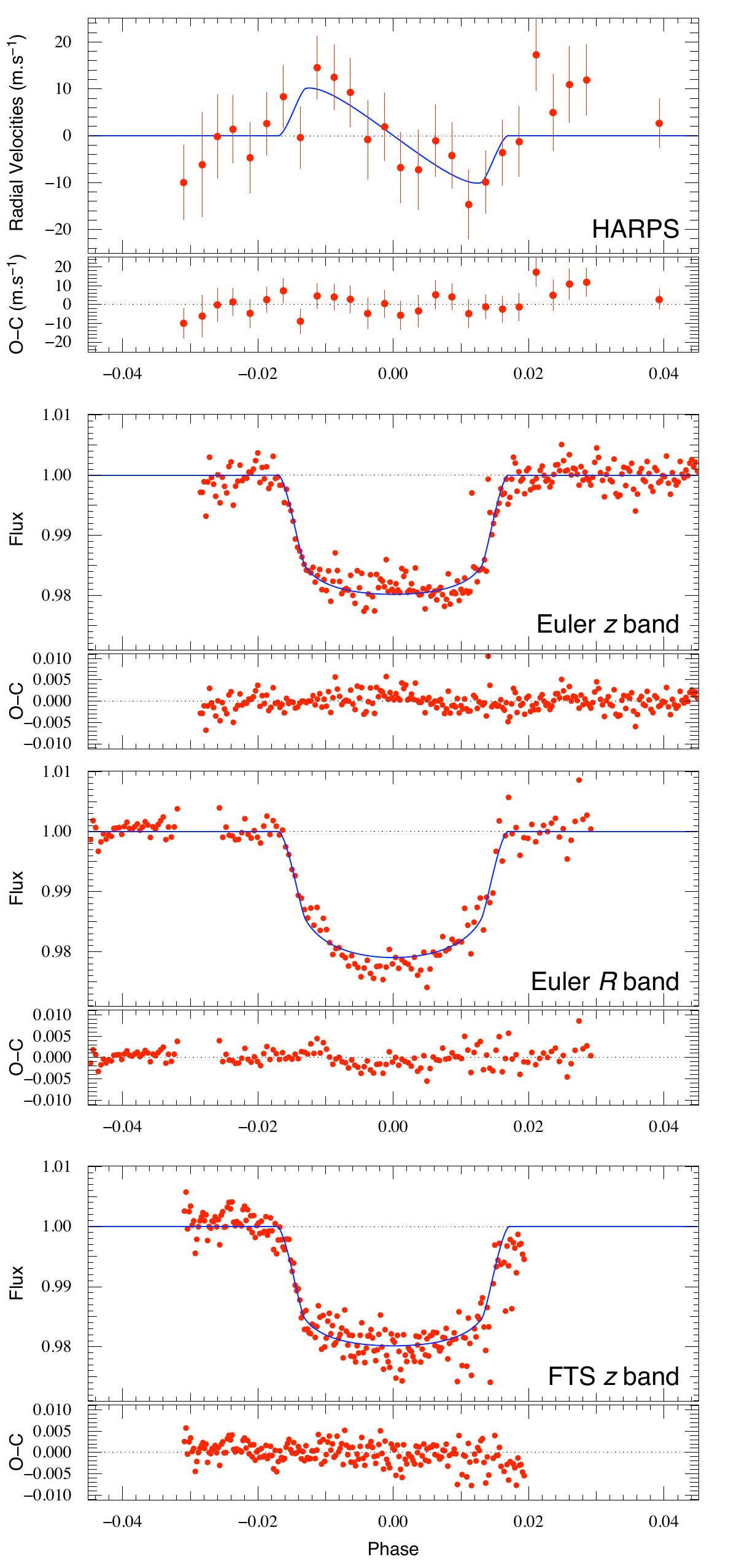}
\caption{HARPS radial velocity data corrected for the reflex Doppler motion due to the planet and model of the Rossiter-McLaughlin effect with residuals. Then three photometric transit as indicated, the best model and residuals.  All are presented as a function of the orbital phase.}\label{fig:euler}

\end{figure}

After a standard pre-reduction, stellar fluxes were extracted using the \textsc{IRAF}\footnote{\textsc{IRAF} is distributed by the National Optical Astronomy Observatory, which is operated by the Association of Universities for Research in Astronomy, Inc., under cooperative agreement with the National Science Foundation} version of the \textsc{DAOPHOT} aperture photometry software \citep{Stetson:1987p2808}. After a careful selection of reference stars, we subtracted a linear fit from the differential magnitudes as a function of airmass to correct for the different colour dependance of the extinction for the target and comparison stars. The linear fit was calculated from the out-of-transit (OOT) data and applied to all the data. The corresponding fluxes were then normalised using the OOT part of the photometry. Fig.~\ref{fig:euler} shows the resulting timeseries. The OOT $rms$ is 2.2 mmag for a mean time sampling of 75s. Comparing this OOT $rms$ to the one obtained after binning the data per 25 minutes (a duration comparable to the one of the ingress/egress) as described by \citet{Gillon:2006p2800} indicates the presence of a correlated noise of $\sim$ 600 ppm in the photometry. 

\subsection{the \textit{Euler} $R$-band transit}\label{subsec:Rtransit}

A similar reduction was operated on the transit of 2010 February 7. After outlier rejection, we were left with 183 images of a mean sampling time of 63s.  In this instance the telescope was not defocused; the mean profile width was 2.3 arcsec. Transparency was good and airmass ranged from 1.03 to 2.22. Five stars were used as reference totalling a comparative flux 4.2 times that of the target. We observe some correlated noise here too, of $\sim$ 600 ppm in the photometry. The photometry is shown on Fig.~\ref{fig:euler}.


\subsection{the FTS $z$-band  transit}\label{subsec:FTS}

An additional transit of WASP-23b was obtained with the LCOGT\footnote{http://lcogt.net} 2.0\,m Faulkes Telescope South (FTS) at Siding Spring Observatory, Australia on the night of 2009 September 27. Observations took place between 15:30 UTC and 19:00 UTC and the airmass decreased throughout from 1.9 at the start to 1.1. The em03 Merope camera was used with a $2\times 2$ binning mode giving a field of view of $5\arcmin \times 5\arcmin$ and a pixel scale of 0.278\,arcsec/pixel. The data were taken through a Pan-STARRS-$z$ filter and the telescope was defocussed to prevent saturation and allow longer 35\,sec exposure times to be used.

The data were pre-processed in the standard manner to perform the debiassing, dark subtraction and flatfielding steps. Aperture photometry was performed using DAOPHOT within the IRAF environment using a 10 pixel radius aperture and the differential photometry was performed relative to 14 comparison stars that were within the FTN field of view.

During the course of the FTS observations, we detected a $\gtrsim 0.55$\,mag deep flat-bottomed partial eclipse on a nearby ($\sim 109$\, arcsec) star (USNO-B1.0 0472-0093932, $\alpha= 06\mathrm{h} 44' 37.63''$ $\delta=-42^{\circ} 45' 13.5''$) appears to be an eclipsing binary. A cursory search of the WASP archive indicates that it is has an ephemeris of HJD(Min I)= 2454021.573377E + 1.421933 days with an eclipse depth of $\sim$\,0.75 mag.
\begin{figure}
\centering                     
\includegraphics[width=9cm]{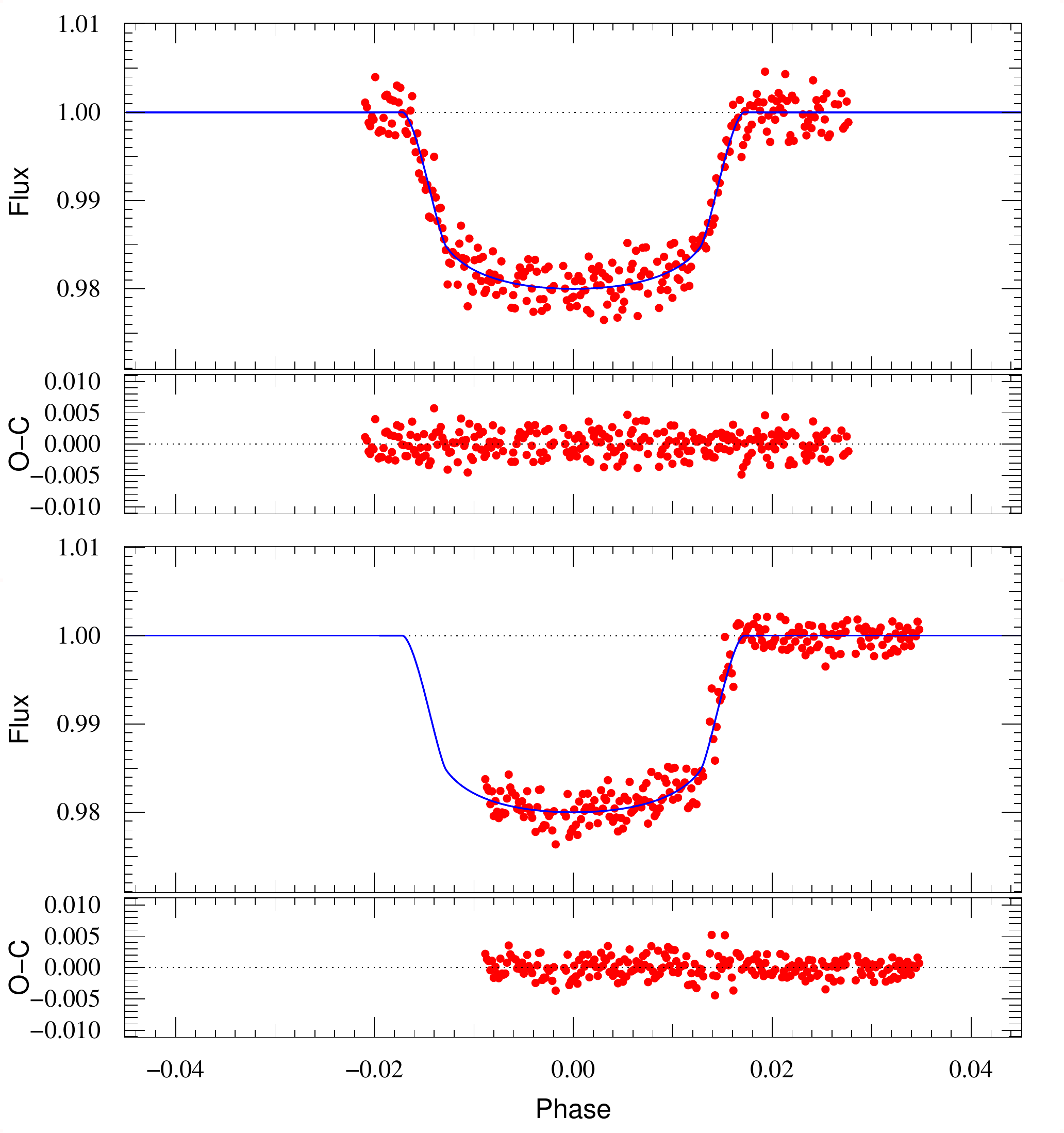}
\caption{The two $I + z$ band timeseries observed by TRAPPIST, with best model overplotted.}\label{fig:trappist}

\end{figure}

\subsection{the TRAPPIST $I+z$ band  transits}\label{subsec:trappist}

A complete and a partial transit  of WASP-23 were also observed  with the robotic 60cm telescope TRAPPIST\footnote{http://arachnos.astro.ulg.ac.be/Sci/Trappist}) \citep{Gillon:2011p8399}. Located at La Silla ESO observatory (Chile), TRAPPIST  is equipped with a 2K $\times$ 2K Fairchild 3041 CCD camera that has a 22' $\times$ 22' field of view (pixel scale = 0.64"/pixel). The transits of WASP-23 were observed on the nights of 2010 December 21 and 30. The sky conditions were clear. We used the 1x2 MHz read-out mode with 1 $\times$ 1 binning, resulting in a typical read-out + overhead time  and read noise of 8.2 s and 13.5 $e^{-}$, respectively. The integration time was 35s for both nights. We observed through a special `I+z' filter that has a transmittance of zero below 700nm, and $>$ 90\% from 750nm to beyond 1100nm. The telescope was  defocused to average pixel-to-pixel sensitivity variations and to optimize the duty cycle, resulting in a typical full width at half-maximum of the stellar images of $\sim$5.2 pixels ($\sim$3.3"). The positions of the stars on the chip were maintained to within a few pixels over the course of the two runs, thanks to the `software guiding' system that derives regularly an astrometric solution on the most recently acquired image and sends pointing corrections to the mount if needed. After a standard pre-reduction (bias, dark, flatfield), the stellar fluxes were extracted from the images using the {\tt IRAF/DAOPHOT}  aperture photometry software \citep{Stetson:1987p2808}. Several sets of reduction parameters were tested, and we kept the one giving the most precise photometry for the stars of brightness similar to WASP-23. After a careful selection of reference stars, differential photometry was obtained.  The data is shows in figure \ref{fig:trappist}.

\subsection{the spectral analysis}\label{subsec:spectral}

A total of 26 individual CORALIE spectra of WASP-23 were co-added to produce a
single spectrum with a typical S/N of around 50:1. The standard pipeline
reduction products were used in the analysis.

The analysis was performed using the methods given in \citet{Gillon:2009p3869}.
The \halpha\ line was used to determine the
effective temperature (\teff), while the Na {\sc i} D and Mg {\sc i} b lines
were used as surface gravity (\logg) diagnostics. The parameters obtained from
the analysis are listed in Table~\ref{wasp23-params}. The elemental abundances
were determined from equivalent width measurements of several clean and
unblended lines. A value for microturbulence (\mictrb) of 0.8 km\,s$^{-1}$\ was determined from
Fe~{\sc i} using \citet{Magain:1984p4690} method. The quoted error estimates include
that given by the uncertainties in \teff, \logg\ and \mictrb, as well as the
scatter due to measurement and atomic data uncertainties.

 The projected stellar rotation velocity (\vsini)\footnote{We make a distinction between $v\,\sin\,I$ and $V\,\sin\,I$. The latter is a result of the Rossiter-McLaughlin fit. $i$ being traditionally the planet's orbital inclination, we denote by $I$, the inclination of the stellar spin axis} was determined by fitting the
profiles of several unblended Fe~{\sc i} lines. Because the value of \vsini was paramount to the model fitting, we used the combined HARPS spectra. A value for macroturbulence
(\mactrb) of $0.8\pm0.3$ km\,s$^{-1}$\  was assumed, based on work by \citet{Bruntt:2010p8010} and an
instrumental FWHM of 0.060\AA, determined from the telluric lines around
6300\AA. A best fitting value of \vsini\ = 2.2 $\pm$ 0.3~km\,s$^{-1}$\ was obtained. Using a macroturbulence based on the tabulation by \citet{Gray:2008p4677} of 1.2 km\,s$^{-1}$\ we obtain the same result for \vsini showing its robustness.

The HARPS spectra show that there is weak emission in the cores of the Calcium H \& K lines. Activity levels on the star are estimated by means of the $\log\,R'_\mathrm{HK}$ \citep{Noyes:1984p6855,Santos:2000p6686,Boisse:2009p1077} and obtained using a $B-V = 0.88 \pm 0.05$ estimated from the effective temperature. The Mount Wilson index, $S_\mathrm{MW}$ is also given.

\subsection{the RV extraction}\label{subsec:RV}

The spectroscopic data were reduced using the online Data Reduction Software (DRS) for the HARPS instrument. The radial velocity information was obtained by removing the instrumental blaze function and cross-correlating each spectrum with a K5 mask. This correlation is compared with the Th-Ar spectrum acting as a reference; see \citet{Baranne:1996p1069} \& \citet{Pepe:2002p1068} for details. Recently the DRS was shown to achieve remarkable precision \citep{Mayor:2009p1088} thanks to a revision of the reference lines for Thorium and Argon by \citet{Lovis:2007p1122}. 
A similar software package is used for CORALIE data. A resolving power $R=110\,000$ for HARPS yields a cross-correlation function (CCF) binned in $0.25$\,km\,s$^{-1}$ increments, while for CORALIE, with a lower resolution of 50\,000, we used $0.5$\,km\,s$^{-1}$. 
The CCF window was adapted to be three times the size of the full width at half maximum (FWHM) of the CCF. 


$1\,\sigma$ error bars on individual data points were estimated from photon noise alone. HARPS is stable on the long term within 1\,m\,s$^{-1}$ and CORALIE to better than 5\,m\,s$^{-1}$. These are smaller than our individual error bars and thus have not been taken into account. 

The absence of a variation in bisector span correlated with the phase, or of a variation of the FWHM, indicate that the photometric and spectroscopic signals are indeed those of a planet. For comparison we invite the reader to look at \citet{Santos:2002p2845}, on HD\,41004 for which it has been proven a blend by a star and its brown dwarf companion was causing a spectroscopic Doppler shift similar to that of a planet on a foreground object. 

\begin{figure}
\centering                     
\includegraphics[width=9cm]{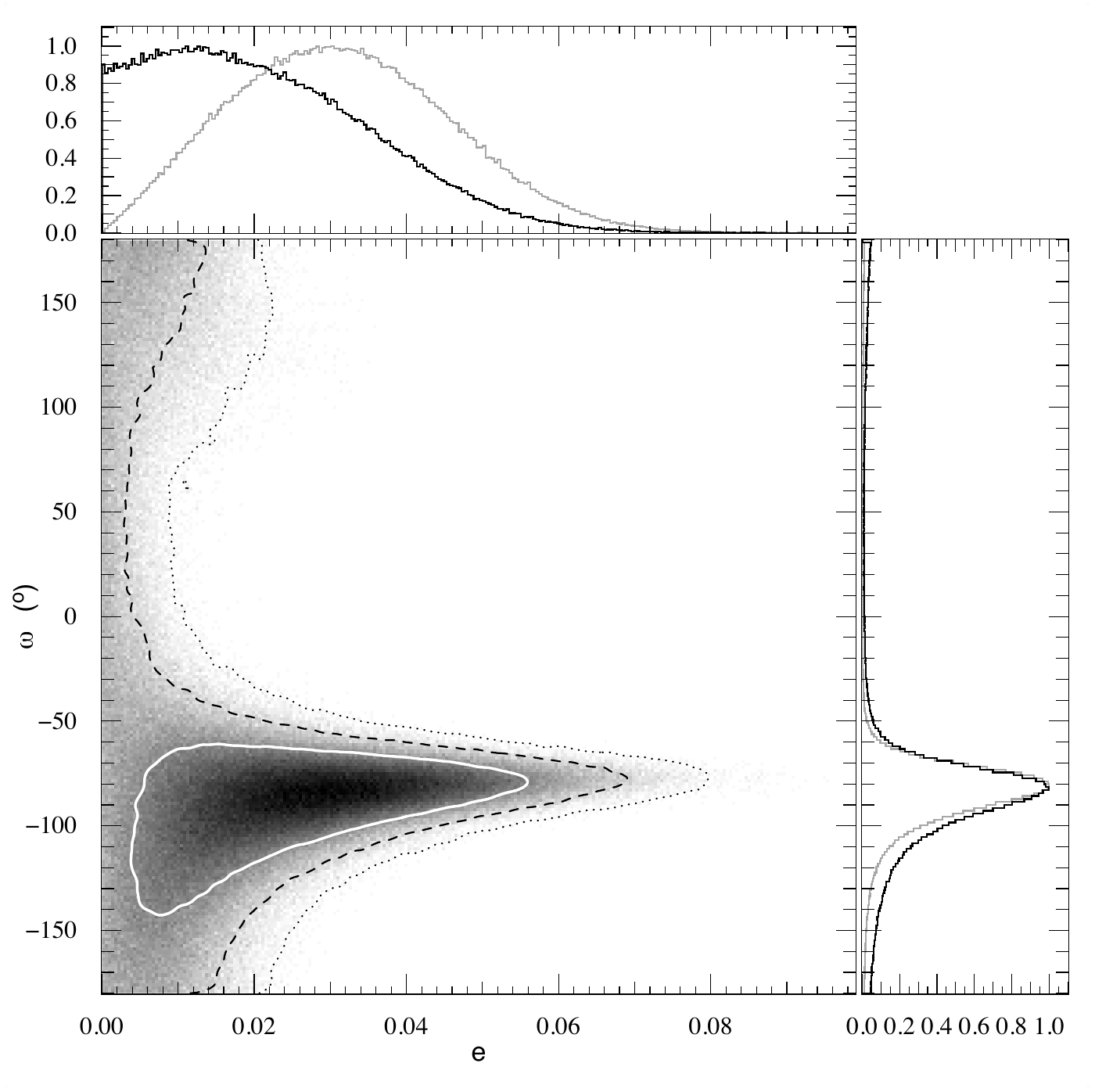}
\caption{In the central box we have the \textit{a posteriori} probability density function for $e$ and $\omega$, resulting from a chain using $\sqrt{e}\,\cos\,\omega$ \& $\sqrt{e}\,\sin\,\omega$ as free parameters (from which $e$ and $\omega$ were computed to fit an eccentric model to the data). The white contour marks the 68.27\,\% confidence region. The black dashed contour shows the 95.45\,\%, and the black dotted contour is the 99.73\,\%. Marginalised distributions are also shown as black histograms in side boxes, normalised to the mode.
Grey histograms in the side boxes show the same fit but instead having $e\,\cos\,\omega$ \& $e\,\sin\,\omega$ as jump parameters. }\label{fig:pdfecc}

\end{figure}

\section{Modelling the data}\label{sec:fitting}

The data was fitted using a Markov Chain Monte-Carlo (MCMC) method. A code allowing to combine both photometry and spectroscopy has been developed. It has been used in several occasions \citep{Bouchy:2008p229, Gillon:2008p767} and is described in length in \citet{Triaud:2009p7520}. It is similar to those presented in \citet{Cameron:2007p2879}. The code uses a common set of free parameters from which physical parameters can be derived to construct models for the photometric and the spectroscopic signal. 

\subsection{parameter choice}

Free parameters used are: $P$ the period of the object, $T_0$ the mid-transit time, $D$ the depth of the transit, $W$ its width, $b$ the impact parameter, $K$ the semi-amplitude of the Doppler reflex motion by the star. To fit the Rossiter-McLaughlin, we use $\sqrt{V\,\sin\,I}\,\cos\,\beta$ and $\sqrt{V\,\sin\,I}\,\sin\,\beta$ where $V\,\sin\,I$ is the projected stellar rotation and $\beta$ the projected spin-orbit angle. Trying to estimate if the orbit is eccentric we used $\sqrt{e}\,\cos\,\omega$ and $\sqrt{e}\,\sin\,\omega$ where $e$ is the eccentricity and $\omega$ is the argument of the periastron. In addition we at times added $\,\dot{\gamma}\,$, a radial-velocity drift with time, in order to assess the presence of an additional body in the system. In addition to these we need to fit one normalisation factor for each photometric dataset (5 in our case) and 2 $\gamma$ velocities for the radial velocities, one for each set. We used Gaussian priors to draw randomly each parameter. 

We decided to use $\sqrt{e}\,\cos\,\omega$ \& $\sqrt{e}\,\sin\,\omega$ as free parameters instead of the more traditional $e\,\cos\,\omega$ \& $e\,\sin\,\omega$ because this would amount to impose a prior proportional to $e^2$ as noted in \citet{Ford:2006p7023}. Figure~\ref{fig:pdfecc} shows the difference between both runs. Having $\sqrt{e}\,\cos\,\omega$ \& $\sqrt{e}\,\sin\,\omega$ makes the eccentricity less biased towards high values. We therefore made a similar change to another pair of variables, making $\sqrt{V\,\sin\,I}\,\cos\,\beta$ and $\sqrt{V\,\sin\,I}\,\sin\,\beta$ as free parameters rather than using
 $V\,\sin\,I\,\cos\,\beta$  and $V\,\sin\,I\,\sin\,\beta$. Checks for these were also conducted and validated our choice of jump parameters.

\begin{figure}
\centering                     
\includegraphics[width=9cm]{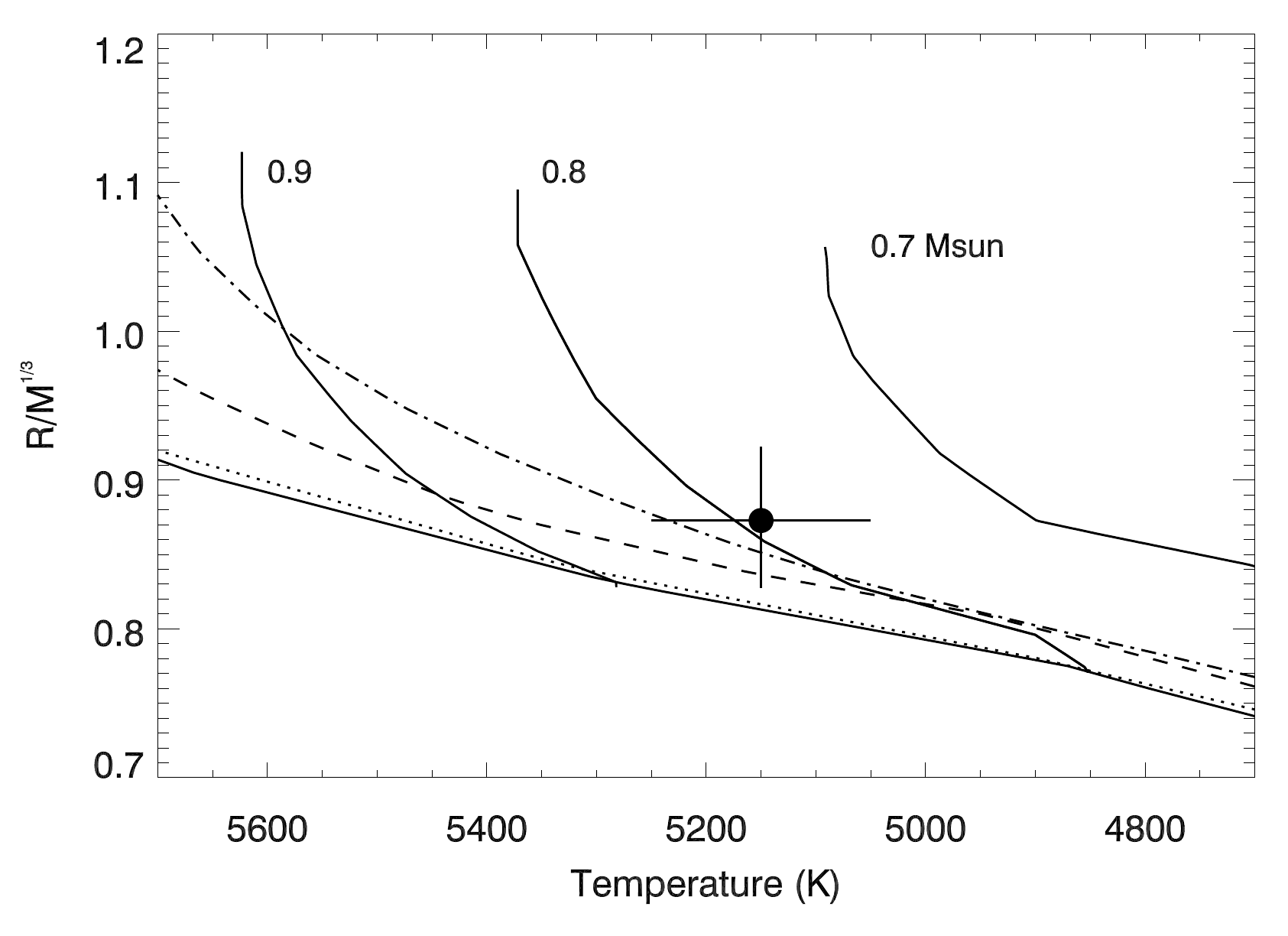}
\caption{Modified Hertzprung-Russell diagram comparing the stellar density and temperature of WASP-23 to theoretical stellar evolutionary tracks interpolated for a metallicity of [M/H] = $-0.05$. The star sits just above the 10\,Gyr tracks (dashdotted line). Other tracks are: 0.1 Gyr (solid), 1 Gyr (dotted), 5 Gyr (dashed). Models are by \citet{Girardi:2000p7510}}\label{fig:evol}

\end{figure}

\subsection{models \& hypotheses}
We used models from \citet{Mandel:2002p768} to fit the photometric transit and from \citet{Gimenez:2006p31} to adjust the Rossiter-McLaughlin effect as well as a classical Keplerian model for the orbital variation in the radial velocities. Limb darkening coefficients for the quadratic law were extracted from \citet{Claret:2000p364,Claret:2004p840} for the photometry. To fit the Rossiter-McLaughlin effect we used coefficients specially made from HARPS's spectral response, that were presented in \citet{Triaud:2009p7520}. Table \ref{tab:limb} shows the values we adopted.

These models are compared to the data using a $\chi^2$ statistics.  A first series of four chains was run and led to a stellar density estimate. This values have been used to determine a stellar mass from evolutionary models by \citet{Girardi:2000p7510} as described in \citet{Hebb:2009p1782} using the metallicity and temperature determined in the spectral analysis and the stellar density from fitting the photometric transit. By interpolating between the tracks we find $M_{\star} = 0.79^{+0.13}_{-0.12}\,M_{\odot}$ (Fig. \ref{fig:evol}). This stellar mass was inserted as a prior in a new series of chains. The stellar age could not be constrained but is likely old; the star sits above the 10 Gyr isochrone.  This first series of chains also allowed the quantification of the correlated noise in the data which is accounted for in the following chains by increasing individual error bars. This allows us to place more credible error bars for parameters determined by the photometry.

\bigskip

Two families of chains, each with 2\,000\,000 random steps, were run. A family consist of four chains with varying hypotheses:
\begin{itemize}
\item eccentricity and RV drift let free;
\item  no eccentricity but RV drift let free;
\item no RV drift but eccentricity let free;
\item no eccentricity and no RV drift .
\end{itemize}

We ran two families, one with the Rossiter-McLaughlin effect let free and one without. Hence from these 8 chains we 
tested for the presence of a linear trend in the radial velocities, for the detection of eccentricity and the detection of the Rossiter-McLaughlin effect. 


\begin{figure}
\centering                     
\includegraphics[width=9cm]{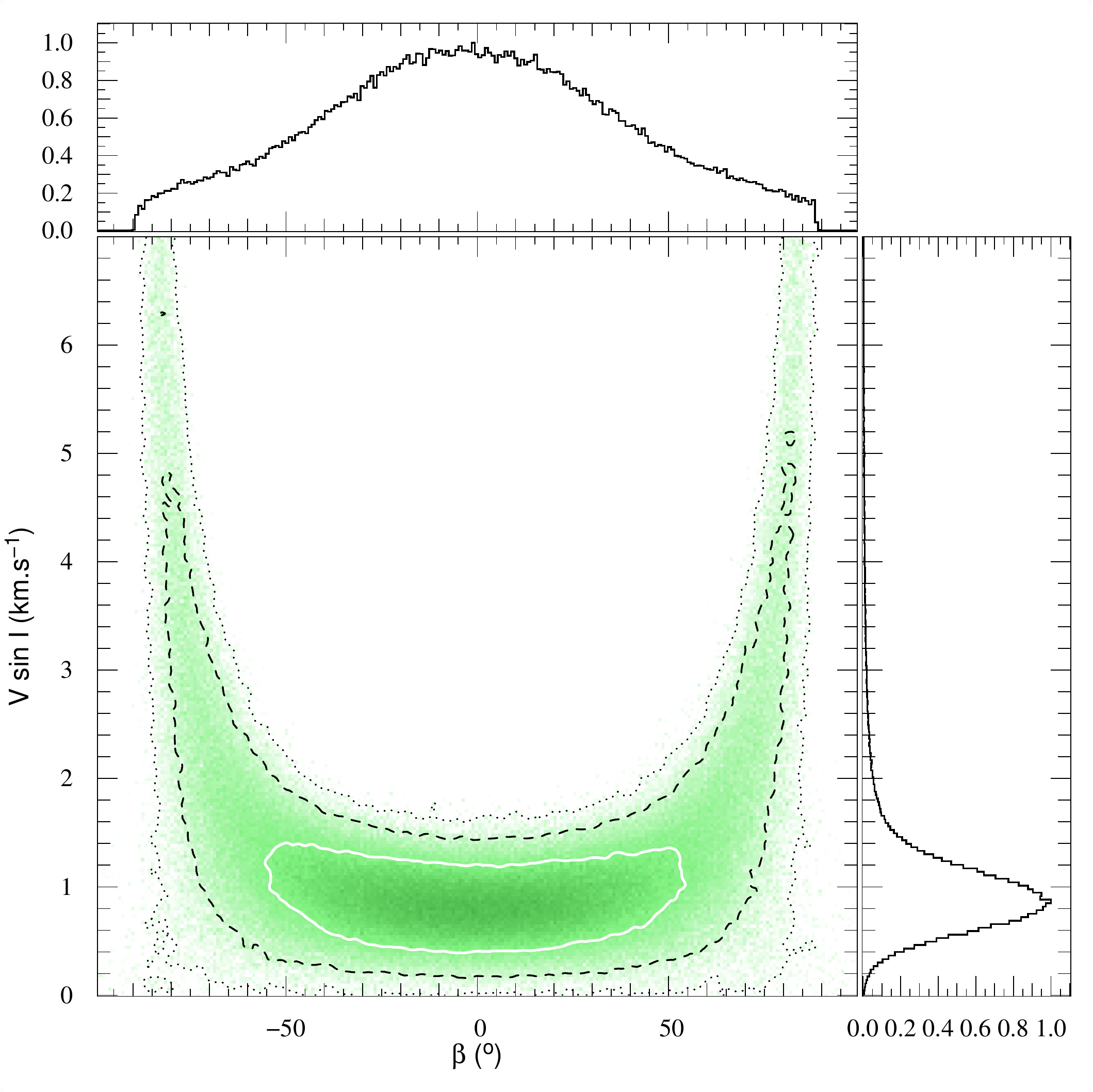}
\caption{Same legend as on Fig.~\ref{fig:pdfecc} but here showing $V\,\sin\,I$ and $\beta$, for a circular, non drifting orbital solution using no prior on $V\,\sin\,I$.  }\label{fig:noprior}

\end{figure}

\section{Results}\label{sec:results}

We have computed 8 different chains, each with different hypotheses. All chains agree in their results within each others' error bars in their common parameters,
giving strong evidence that indeed they have converged to the solution. Results were extracted from each chain and their comparison with each others lead to the final results here presented. This comparison, available for its most interesting parts in the appendices, made us choose a circular, non drifting model for the radial velocity. WASP-23b's parameters are extracted by taking the mode of the marginalised distributions computed by the Markov chains. When two clear separated mode appear, each is estimated and errors bars calculated (see Fig.~\ref{fig:pdf}). Error bars are computed by taking the 68.27\%, 95.45\% and 99.73\% marginalised confidence regions in the \textit{a posteriori} probability density distribution.  Models using eccentricity and a drift as free floating parameters are useful to place upper constraints on these parameters.  Results are presented in Table~\ref{tab:params}. The final reduced $\chi^2$ for the radial velocities, both for CORALIE and HARPS observations, is consistent with 1. We therefore saw no need to add any additional terms to our error bars.

WASP-23b, therefore, is a $0.88\,M_\mathrm{J}$ planet with a $0.96\,R_\mathrm{J}$ radius on a 2.94 day orbit, placing it among the normally sized hot Jupiter planets. We can reasonably estimate upper limits on the eccentricity and RV trend  at the 99\% exclusion level, from the chains that had these as free parameters. Thus we find that $e < 0.062$ \& $|\,\dot{\gamma}\,| < 30$ m\,s$^{-1}$\,yr$^{-1}$. 


Using no prior, we have detected a radial velocity anomaly compatible with the expected location and shape of the Rossiter-McLaughlin ($\chi^2$ changes from $19.6\pm6.2$ to $6.4\pm3.6$ for the 14 data points positioned during transit and one on either side). We detected this effect to a confidence of $3.2\,\sigma$ from a simple Keplerian model. We see a degeneracy arising between $\beta$, $V\,\,sin\,I$ \& $b$, the impact parameter: $V\,\sin\,I$ grows to extremely large values (up to 60 km\,s$^{-1}$, Fig.~\ref{fig:noprior}) by forcing $\beta$ to severely misaligned solution and $b$ closer to an equatorial transit. 

Because of the non physicality of such result for $V\,\sin\,I$ we will have to resort to the use of some prior in order to constrain the space within which the MCMC can explore. We do this by imposing a Bayesian penalty on $\chi^2$. This additional analysis is presented in the discussion.

Our chains indicate that the effect is mostly symmetrical with respect to the centre of transit (see Fig. \ref{fig:noprior}), and that we can exclude a retrograde orbit.  


\begin{table*}
\caption{Fitted and derived parameters for WASP-23 \& WASP-23b with their error bars for confidence intervals of 68.27\%, 95.45\% and 99.73\%. For asterisked parameters, please refer to the text. Underscripted 1 and 2 are for  showing 2 distinct solutions for the same parameter}\label{tab:params}
\begin{tabular}{lllll}
\hline
\hline
Parameters $(units)$ && $1\,\sigma$&$2\,\sigma$&$3\,\sigma$\\
\hline
&\\
\multicolumn{2}{l}{\textit{fitted parameters}} &\\
&\\
$P$ (days)                              & $2.9444256$    &$^{+0.0000011}_{-0.0000013}$ &$^{+0.0000024}_{-0.0000024}$ &$^{+0.0000036}_{-0.0000036}$ \\
$T_0$ (bjd-2\,450\,000)       & $5320.12363$  &$^{+0.00012}_{-0.00013}$          &$^{+0.00023}_{-0.00026}$          &$^{+0.00036}_{-0.00039}$      \\
$D$                                          &$ 0.01691$         &$^{+0.00010}_{-0.00011}$          &$^{+0.00024}_{-0.00024}$          &$^{+0.00053}_{-0.00034}$   \\
$W$ (days)                             &$ 0.09976 $        &$^{+0.00031}_{-0.00039}$          &$^{+0.00081}_{-0.00077}$          &$^{+0.00188}_{-0.00116}$       \\
$b_1$ ($R_{\star}$) *                 &$ 0.04           $     &$^{+0.05}_{-0.04}$                        &$^{+0.17}_{-0.04}$                        &$^{+0.33}_{-0.04}$                \\
$b_2$ ($R_{\star}$) *                 &$ 0.05           $     &$^{+0.23}_{-0.05}$                        &$^{+0.31}_{-0.05}$                        &$^{+0.37}_{-0.05}$                \\
$K$ (m\,s$^{-1}$)                   &$ 145.8       $      &$^{+1.5}_{-2.1}$                             &$^{+3.4}_{-4.0}$                             &$^{+5.3}_{-5.6}$                  \\
$\sqrt{V\,\sin\,I}_1\,\cos\,\beta_1$ *& $ 0.57 $             &$^{+0.18}_{-0.16}$                         &$^{+0.42}_{-0.34}$                       &$^{+0.66}_{-0.46}$\\
$\sqrt{V\,\sin\,I}_1\,\sin\,\beta_1$  *& $ -1.4 $              &$^{+2.8}_{-0.1}$                              &$^{+3.0}_{-0.3}$                           &$^{+3.0}_{-0.4}$\\
$\sqrt{V\,\sin\,I}_1\,\cos\,\beta_2$ *& $ 1.00 $             &$^{+0.09}_{-0.29}$                         &$^{+0.16}_{-0.56}$                       &$^{+0.23}_{-0.79}$\\
$\sqrt{V\,\sin\,I}_1\,\sin\,\beta_2$  *& $ -0.9 $              &$^{+1.9}_{-0.2}$                              &$^{+1.9}_{-0.2}$                           &$^{+2.1}_{-0.4}$\\
      &        \\
      &        \\
\multicolumn{2}{l}{\textit{derived parameters} }      &      \\
&\\
$R_p / R_{\star}$                         & $ 0.13004 $  &$^{+0.00040}_{-0.00045}$  &$^{+0.00095}_{-0.00091}$  &$^{+0.00203}_{-0.00132}$ \\
 & \\
$R_{\star} / a$                               & $ 0.09429$  &$^{+0.00041}_{-0.00047}$  &$^{+0.00212}_{-0.00091}$  &$^{+0.00675}_{-0.00124}$\\
$\rho_{\star}$   $(\rho_{\odot})$ & $ 1.843$       &$^{+0.025}_{-0.027}$    &$^{+0.054}_{-0.119}$    &$^{+0.069}_{-0.347}$\\
$R_{\star}$  $(R_{\odot})$          & $0.765$      &$^{+0.033}_{-0.049}$      &$^{+0.068}_{-0.098}$       &$^{+0.102}_{-0.164}$          \\
$M_{\star} $  $(M_{\odot})$        & $ 0.78$       &$^{+0.13}_{-0.12}$  \\
 & \\
$R_p / a$                                       & $ 0.012260$      &$^{+0.000077}_{-0.000077}$  &$^{+0.000340}_{-0.000168}$  &$^{+0.001093}_{-0.000222}$\\
$R_p$  $(R_\mathrm{J}$)            & $ 0.962$         &$^{+0.047}_{-0.056}$           &$^{+0.095}_{-0.118}$           &$^{+0.139}_{-0.199}$\\
$M_p$ $(M_\mathrm{J})$           & $0.884$             &$^{+0.088}_{-0.099}$                 &$^{+0.178}_{-0.203}$               &$^{+0.262}_{-0.321}$   \\
&     \\
$a$ (AU)                            & $ 0.0376 $              &$^{+0.0016}_{-0.0024}$  &$^{+0.0034}_{-0.0046}$  &$^{+0.0049}_{-0.0078}$  \\
$i$   $(^{\circ})$                 & $ 88.39 $                   &$^{+0.79}_{-0.45}$        &$^{+1.50}_{-0.69}$            &$^{+1.56}_{-1.03}$ \\
\\
$V\,\sin\,I_1$ (km\,s$^{-1}$) *& $ 2.03 $                   &$^{+0.37}_{-0.35}$        &$^{+0.70}_{-0.70}$            &$^{+0.99}_{-1.00}$ \\
$|\, \beta_1\,|$   ($^{\circ}$)  * & $ 69$                      &$^{+6}_{-9}$                     &$^{+14}_{-24}$                  &$^{+18}_{-65}$ \\
$V\,\sin\,I_2$ (km\,s$^{-1}$) *& $ 1.21 $                   &$^{+0.17}_{-0.23}$        &$^{+0.42}_{-0.39}$            &$^{+0.64}_{-0.52}$ \\
$\beta_2$   ($^{\circ}$)         * & $ -43$                     &$^{+99}_{-17}$                 &$^{+109}_{-22}$              &$^{+122}_{-35}$ \\
\\
$e$                                      & &&&$  < 0.062 $                \\
$|\,\dot{\gamma}\,|$     (m\,s$^{-1}$\,yr$^{-1}$)                &&        &         & $  < 30 $                \\\\
\\
\multicolumn{2}{l}{$\gamma$ \textit{velocity} (m\,s$^{-1}$)}\\
\\
CORALIE&$5674.403 $&$^{+0.040}_{-0.046}$        &$^{+0.085}_{-0.088}$            &$^{+0.130}_{-0.136}$\\
HARPS&$5691.60$&$^{+0.33}_{-0.84}$        &$^{+0.90}_{-1.38}$            &$^{+1.44}_{-1.97}$\\
\\
\multicolumn{2}{l}{\textit{normalisation factors}}\\
\\
WASP-South & $1.00068$             &$^{+0.000011}_{-0.000011}$        &$^{+0.000021}_{-0.000022}$            &$^{+0.000032}_{-0.000033}$\\
-& $1.000272$&$^{+0.0000020}_{-0.0000021}$        &$^{+0.0000041}_{-0.0000041}$            &$^{+0.0000060}_{-0.0000062}$\\
\textit{Euler z}-band &$1.00013$ &$^{+0.000052}_{-0.000086}$        &$^{+0.000125}_{-0.000159}$            &$^{+0.000190}_{-0.000232}$\\
\textit{Euler R}-band &$0.99998$ &$^{+0.000064}_{-0.000052}$        &$^{+0.000122}_{-0.000112}$            &$^{+0.000178}_{-0.000176}$\\
\textit{FTS z}-band &$1.01110$    &$^{+0.00011}_{-0.00010}$            &$^{+0.00022}_{-0.00021}$                &$^{+0.00032}_{-0.00033}$\\
\textit{TRAPPIST I + z}-band &$1.000117$    &$^{+0.000078}_{-0.000078}$            &$^{+0.000163}_{-0.000160}$                &$^{+0.000239}_{-0.000239}$\\
 &$1.000150$    &$^{+0.000059}_{-0.000068}$            &$^{+0.000125}_{-0.000130}$                &$^{+0.000192}_{-0.000195}$\\

\hline

\end{tabular}

\end{table*}

\section{Discussion}

\begin{figure*}
\centering                     
\includegraphics[width=16.5cm,angle=0]{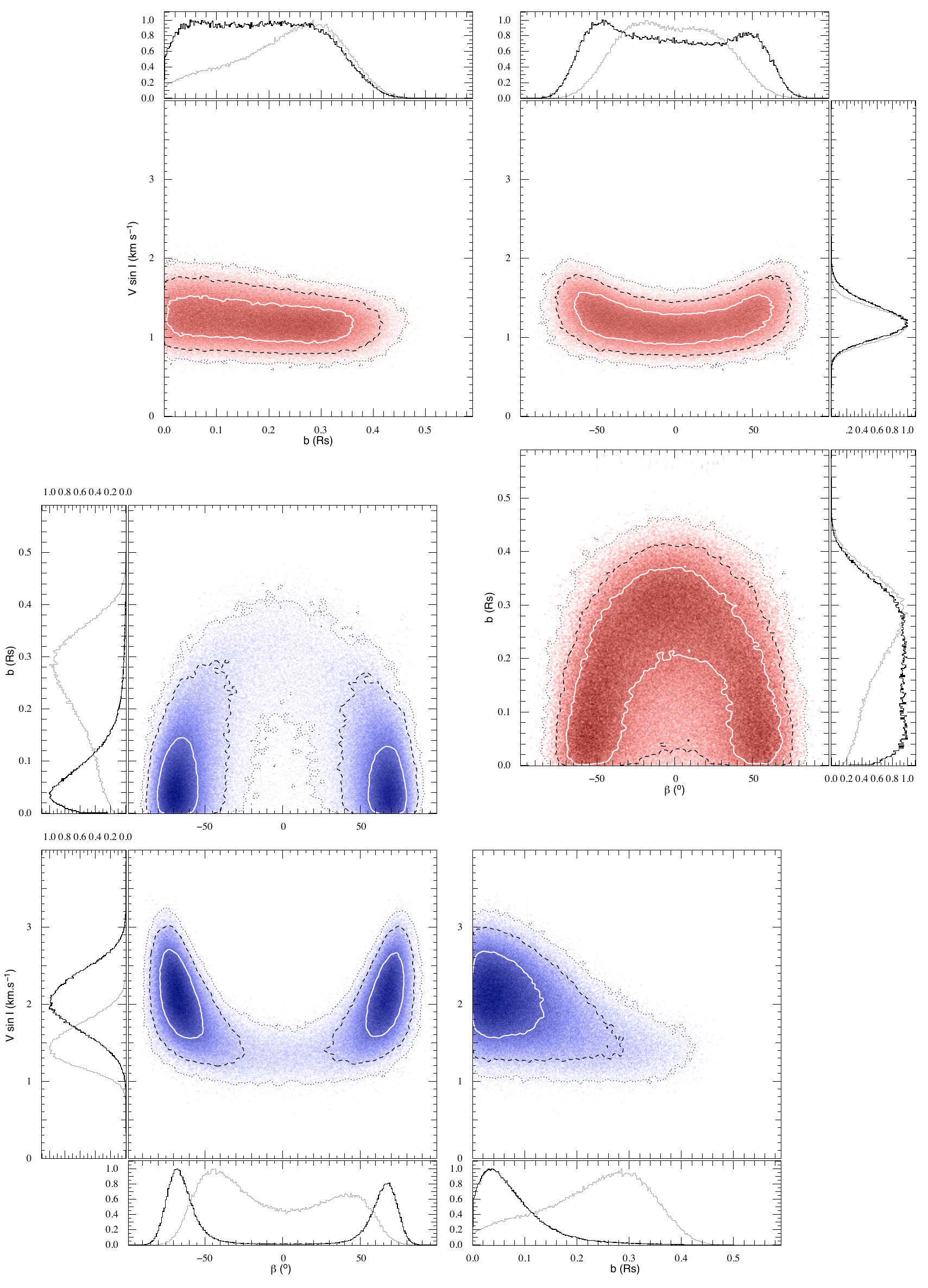}
\caption{Square boxes present the \textit{a posteriori} probability density functions for $V\,\sin\,I$, $\beta$ and impact parameter $b$, from which we extract our results.  The white contour marks the 68.27\,\% confidence region. The black dashed contour shows the 95.45\,\%, and the black dotted contour is the 99.73\,\%. Marginalised distributions are also shown as black histograms in side boxes, normalised to the mode. On top right, in red, we have the results for a circular, non drifting solution with use of a prior of $v  = 1.35 \pm 0.20$ km\,s$^{-1}$. On bottom left, in blue, we show a circular, non drifting solution with the application of a prior of $v\,\sin\,I = 2.2 \pm 0.3$ km\,s$^{-1}$. Grey histograms in $V\,\sin\,I$ and $\beta$ show results in the photometry limited runs; for $b$ we plotted the resulting distribution by fitting the photometry alone without the Rossiter-McLaughlin effect. }\label{fig:pdf}

\end{figure*}

Figure~\ref{fig:noprior} shows unambiguously that we have a strong degeneracy between $\beta$ and $V\,\sin\,I$. This is a known problem notably reported already in \citet{Narita:2010p8435} \& \citet{Triaud:2010p8039}. Spectral line analysis (section \ref{subsec:spectral}) showed that the projected stellar rotation velocity \vsini  = 2.2 $\pm$ 0.3 km\,s$^{-1}$. From figure~\ref{fig:noprior} we observe that a $V\,\sin\,I$ at such a value would lead to a severely misaligned orbit. This is confirmed when running the MCMC for an additional family of chains using this value of \vsini as a prior on $V\,\sin\,I$. 

We call this our solution 1:
we obtain two modes for $\beta$ symmetrically opposite to each other (see Fig. \ref{fig:pdf}, bottom left corner). We therefore choose to denote this result by its absolute value as $|\, \beta_1\,| = 69^{\circ}$$^{+6}_{-9}$; $V\,\sin\,I_1 =  2.03^{+0.37}_{-0.35}\,$ km\,s$^{-1}$. Impact parameter $b_1 = 0.04 \pm 0.05\,R_{\star}$. This prior choice makes the detection of the Rossiter-McLaughlin effect close to $7.5\,\sigma$.

Before claiming an additional misaligned planet, one around a cool star (thus a strong exception to  \citet{Winn:2010p7311}), we have searched for another independent estimate of stellar rotation.

From spectral analysis, we obtained a value quantifying emission in the Calcium II lines expressed in the form of $\log\, R'_\mathrm{HK} = -4.68\pm0.07$. This value can be used as an indirect measurement of the true stellar rotation period. We used two methods, by \citet{Noyes:1984p6855}, finding $28.6^{+5.3}_{-5.3}$ days and a more recent estimate by \citet{Mamajek:2008p8051} and got the value $28.2^{+4.4}_{-5.3}$ days in very good agreement with the previous. Using the distribution of $R_{\star}$ from our MCMC chains, we transformed these values in the equatorial rotation $v$ obtaining $1.30^{+0.24}_{-0.19}$ and $1.35^{+0.28}_{-0.20}$ km\,s$^{-1}$ respectively. Combining both we have our new prior 
that we included as $1.35 \pm 0.20$ km\,s$^{-1}$ into a new series of four chains. Results from these indicate that the planet is most likely on an aligned orbit, but error bars remain large. 

We call these results, our solution 2:
we only have one large range of values for the projected spin-orbit angle: $\beta_2=-43^{\circ}$$^{+99}_{-17}$. $V\,\sin\,I_2 = 1.21^{+0.17}_{-0.23}$ km\,s$^{-1}$ and $b_2 = 0.05^{+0.23}_{-0.05}\, R_{\star}$. Here too, the detection of the Rossiter-McLaughlin is more secure, climbing to a value close to $7\,\sigma$.
\\



In order to test our results we simulated whether an infinitely precise photometry would help discriminate between our differing solutions for $\beta$ since the parameter $b$, for example, is influenced by the adjustment of the Rossiter-McLaughlin model. We therefore removed the Rossiter-McLaughlin effect and ran a chain only using photometry on the transit parameters sticking with the non drifting, non eccentric model.  Having photometry as the sole influence on the impact parameter, one finds $b = 0.28^{+0.08}_{-0.14}\, R_{\star}$. The value found for $b_1$ and $b_2$ are not at odds with this value. We fixed the parameters controlled by photometry to the best found values (ie. $b = 0.28$) and ran an additional three chains. Those three "photometry limited" chains had the following differences: 
\begin{itemize}
\item no imposed prior;
\item a prior of 2.2 $\pm$ 0.3 km\,s$^{-1}$; 
\item a prior of $1.35 \pm 0.20$ km\,s$^{-1}$.
\end{itemize}

Refer to figure~\ref{fig:pdf} to see the resulting probability distributions (grey histograms). By fixing all parameters to the values that can be determined by photometry alone and letting free again all remaining parameters, we see some changes in the posterior probability distributions (see grey histograms on Fig. \ref{fig:pdf}). The distribution of $V\,\sin\,I_1$ is shifted to lower values, while that of $\beta_1$, although still bimodal, is less so. $V\,\sin\,I_2$ is left almost unchanged, but $\beta_2$ is more confined around 0. This shows what effect our poorly constrained impact parameter induces on our fit, but this also indicates that, unless a new transit finds a higher value of $b$, even an infinitely precise photometry would not help lifting entirely the degeneracy between $\beta$ and $V\,\sin\,I$ for $b < 0.28 \, R_{\star}$.

The measurement of the projected spin-orbit angle $\beta$ is mostly affected by the low amplitude of the signal and in part by the poorly determined impact parameter, which, floating to small values, helps create a degeneracy between small $\beta$, small $V\,\sin\,I$ and high $\beta$, high $V\,\sin\,I$.

\section{Conclusions}

After the analysis of more than 4 years of photometric and spectroscopic data, we confidently conclude that we have detected a typical hot Jupiter around the K1V star called WASP-23. The analysis was made a bit more arduous because of degeneracy arising when adjusting for the Rossiter-McLaughlin effect. A total of 19 Markov chains were used to derive our conclusions.
 
Despite the slow rotation and likely old age of the star, we manage to detect the Rossiter-McLaughlin effect, which is of a similar amplitude to that of WASP-2b \citep{Triaud:2010p8039} and of a similar signal to noise to detections like that on Hat-P-11b \citep{Winn:2010p8446,Hirano:2010p8601}. Imposing priors on $V\,\sin\,I$ increases our detection level but seriously affects the posterior probability distribution (see Fig. \ref{fig:noprior} \& \ref{fig:pdf}) and thus our results and our possible interpretations. This mostly come from the difference between our two priors, which are $2.7\,\sigma$ away from each other although our value of $v\,\sin\,I = 2.2 \pm 0.3$ km\,s$^{-1}$ (solution 1) should be seen as a lower value, while $v  = 1.35 \pm 0.20$ km\,s$^{-1}$ (solution 2) is more of an upper value.
 
There is strong evidence that for stars colder than 6250\,K, planets tend to have a high \textit{a priori} probability to be aligned with the stellar spin \citep{Winn:2010p7311}. This then would support our solution 2 and thus imply that the spectral line broadening method is not accurate enough to determine $v\,\sin\,I$. This would demonstrate that there is a potential difficulty with the estimation of $v\,\sin\,I$ and its use as a prior. 
But solution 2 is not without its problems either since the determination of the stellar rotation from activity indices could be altered by the presence of a nearby hot Jupiter and long term magnetic cycles. 

To resolve WASP-23b's spin orbit angle, one has several options. 
One could acquire more and better photometry. If $b$ were to be found small, then the degeneracy will not be lifted, but the larger it is found, the more likely the system will be aligned. Additionally, further observations might be needed of the Rossiter-McLaughlin. There is also a possibility that the current data is enough as one could use the \textit{Doppler shadow} method pioneered in \citet{CollierCameron:2010p5889} and \citet{Miller:2010p8044}. This method provides precise determinations of $V\,\sin\,I$, $\beta$ and $b$ but works best for fast rotators and bright targets. 

Either way, the resolution of our degenerate result is interesting. If solution 1 (prior on $V\,\sin\,I =  2.2 \pm 0.3$ km\,s$^{-1}$) is confirmed, we have a misaligned system around a cold star and a need to rectify the determination of stellar rotation from activity levels. If solution 2 (prior on $V\,\sin\,I =  1.35 \pm 0.20$ km\,s$^{-1}$) is preferred, then we have probably and aligned system and one will know to be careful when using $v\,\sin\,I$ prior found by spectral line broadening.

We therefore recommend extreme caution when using priors as final results can entirely depend on those and a small initial systematic error can lead to dramatic changes in interpretation.

\paragraph{\textbf{Nota Bene}}
We used the UTC time standard and Barycentric Julian Dates in our analysis. Our results are using the equatorial solar and jovian radii and masses from Allen's Astrophysical Quantities.

\begin{acknowledgements} 
The authors would like to acknowledge the use of ADS and of \textit{Simbad} but mostly the help and the kind attention of the ESO staff at La Silla which with its (now gone and much regretted) \textit{completos} insured a good and stable supply of energy during the long observing nights. The work is supported by the Swiss Fond National de Recherche Scientifique. 
TRAPPIST is a project funded by the Belgian Fund for Scientific Research (FNRS) with the participation of the Swiss National Science Fundation (SNF). M. Gillon and E. Jehin are FNRS Research Associate. 
We also thank R. Heller for providing a copy of Holt's 1893 paper.

\end{acknowledgements} 

\bibliographystyle{aa}
\bibliography{bibtex}
\appendix

\section{Comparisons between different chains.}





\begin{table*}
\caption{Here we compare the results in $\chi^2$ by instrument and during the Rossiter-McLaughlin effect 2 different families of chains: having no prior on $V\,\sin\,I$, and not fitting the Rossiter-McLaughlin effect.}\label{tab:comp}
\begin{tabular}{lllll}
\hline
\hline
($e$, $|\,\dot{\gamma}\,|$) & (free, free) &  (free,fixed) & (fixed, free) & (fixed, fixed)\\
\hline
\\
\multicolumn{4}{l}{\textit{Rossiter-McLaughlin effect fitted, no Prior}}\\ 
\\
$\chi^2_\mathrm{CORALIE,\,38\,RVs}$      &$38.1\pm8.7$         &$45.9\pm9.6$          &$41.3\pm9.1$      &$48.5\pm9.8$   \\
$N_\mathrm{param}$                                     &7                                &6                                & 5                            & 4                         \\
$\chi^2_\mathrm{reduced}$                          &$1.23\pm0.28$       &$1.44\pm0.30$       &$1.25\pm0.28$    &$1.43\pm0.29$  \\
\\
$\chi^2_\mathrm{HARPS,\,35\,RVs}$          &$25.2\pm7.1$         &$23.8\pm7.0$          &$27.1\pm7.4$      &$26.2\pm7.2$     \\
$N_\mathrm{param}$                                     &12                              &11                              & 10                          & 9                          \\
$\chi^2_\mathrm{reduced}$                          &$1.10\pm0.31$       &$1.00\pm0.29$        &$1.08\pm0.29$    &$1.01 \pm0.28$  \\
\\
$\chi^2_\mathrm{in\,RM,\,16\,RVs}$             &$6.8\pm3.7$         &$6.8\pm3.7$          &$6.3\pm3.6$      &$6.4\pm3.6$     \\
\\
\multicolumn{2}{l}{\textit{all 73 RVs, 2 sets}}\\
$\chi^2_\mathrm{RV}$                                    &$63.4\pm11.3$       &$69.7\pm11.8$        &$68.4\pm11.7$    &$74.7\pm12.2$   \\
$N_\mathrm{param}$                                      &13                             &12                               & 11                         & 10                          \\
$\chi^2_\mathrm{reduced}$                           &$1.06\pm0.19$       &$1.14\pm0.19$        &$1.10\pm0.19$    &$1.19\pm0.19$   \\
\\
\hline
\\
\multicolumn{4}{l}{\textit{Rossiter-McLaughlin effect absent}}\\
\\
$\chi^2_\mathrm{CORALIE,\,38\,RVs}$      &$38.1\pm8.7$         &$45.1\pm9.5$          &$40.7\pm9.0$      &$48.4\pm9.8$   \\
$N_\mathrm{param}$                                     &7                                &6                                & 5                            & 4                         \\
$\chi^2_\mathrm{reduced}$                          &$1.23\pm0.28$       &$1.41\pm0.30$       &$1.23\pm0.27$    &$1.42\pm0.29$  \\
\\
$\chi^2_\mathrm{HARPS,\,35\,RVs}$          &$39.7\pm8.9$         &$38.9\pm8.8$          &$40.8\pm9.0$      &$39.5\pm8.9$     \\
$N_\mathrm{param}$                                     &7                                &6                                & 5                            & 4                          \\
$\chi^2_\mathrm{reduced}$                          &$1.42\pm0.32$       &$1.34\pm0.30$        &$1.36\pm0.30$    &$1.27 \pm0.29$  \\
\\
$\chi^2_\mathrm{in\,RM,\,16\,RVs}$                &$21.2\pm6.5$         &$21.2\pm6.5$          &$19.6\pm6.3$      &$19.6\pm6.2$     \\
\\
\multicolumn{2}{l}{\textit{all 73 RVs, 2 sets}}\\
$\chi^2_\mathrm{RV}$                                    &$77.9\pm12.5$       &$84.1\pm13.0$        &$81.5\pm12.8$    &$87.9\pm13.3$   \\
$N_\mathrm{param}$                                      &8                                &7                                 & 6                            & 5                          \\
$\chi^2_\mathrm{reduced}$                           &$1.20\pm0.19$       &$1.27\pm0.20$        &$1.22\pm0.19$    &$1.29\pm0.20$   \\
\\
\hline

\end{tabular}
\end{table*}

\begin{table*}
\caption{Here are presented the results from various Markov chains of the three parameters which control the shape of the Rossiter-McLaughlin effect. These are for circular non drifting orbital solutions.}\label{tab:degen}
\begin{tabular}{lllllllllllllll}
\hline
&\multicolumn{4}{l}{$V\,\sin\,I$ (km\,s$^{-1}$) $\pm$}&\multicolumn{4}{l}{$\beta$ ($^{\circ}$) $\pm$}&&\multicolumn{4}{l}{$b$ ($R_{\star}$) $\pm$}\\
&&$1\,\sigma$&$2\,\sigma$&$3\,\sigma$&&&$1\,\sigma$&$2\,\sigma$&$3\,\sigma$&&&$1\,\sigma$&$2\,\sigma$&$3\,\sigma$\\
\hline
\\
\textit{$V\,\sin\,I$ Prior off} \\
&0.82&$^{+0.11}_{-0.35}$&$^{+1.98}_{-0.82}$&$^{+9.45}_{-0.82}$
&&-1&$^{+38}_{-41}$&$^{+76}_{-76}$&$^{+89}_{-86}$
&&0.27&$^{+0.05}_{-0.27}$&$^{+0.10}_{-0.27}$&$^{+0.15}_{-0.27}$\\
\\
\textit{$V\,\sin\,I$ Prior on, 1.35 km\,s$^{-1}$} \\
&1.21&$^{+0.17}_{-0.23}$&$^{+0.42}_{-0.39}$&$^{+0.64}_{-0.52}$
&&-43&$^{+99}_{-17}$&$^{+109}_{-22}$&$^{+122}_{-35}$
&&0.05&$^{+0.23}_{-0.02}$&$^{+0.31}_{-0.05}$&$^{+0.37}_{-0.05}$\\
\\
\textit{$V\,\sin\,I$ Prior on, 2.2 km\,s$^{-1}$} \\
&2.03&$^{+0.37}_{-0.35}$&$^{+0.70}_{-0.70}$&$^{+0.99}_{-1.00}$
&& \,\,68&$^{+7}_{-8}$&$^{+14}_{-23}$&$^{+18}_{-155}$
&&0.04&$^{+0.05}_{-0.04}$&$^{+0.17}_{-0.04}$&$^{+0.33}_{-0.04}$\\
&&&&
&&-69&$^{+8}_{-7}$&$^{+25}_{-14}$&$^{+155}_{-18}$\\
\\
\\
\multicolumn{5}{l}{Rossiter-McLaughlin effect not fitted }\\
\hline
\\
&0&\multicolumn{3}{c}{(fixed)}
&&0&\multicolumn{3}{c}{(fixed)}
&&0.28&$^{+0.08}_{-0.14}$&$^{+0.11}_{-0.26}$&$^{+0.15}_{-0.28}$\\
\\

\multicolumn{3}{l}{Photometry Limited runs}\\
\hline
\\
\textit{$V\,\sin\,I$ Prior off} \\
&0.87&$^{+0.28}_{-0.28}$&$^{+0.57}_{-0.53}$&$^{+0.87}_{-0.80}$
&& 4&$^{+23}_{-36}$&$^{+46}_{-57}$&$^{+70}_{-77}$
&&0.28&\multicolumn{3}{c}{(fixed)}\\
\\
\textit{$V\,\sin\,I$ Prior on, 1.35 km\,s$^{-1}$} \\
&1.15&$^{+0.19}_{-0.17}$&$^{+0.37}_{-0.33}$&$^{+0.54}_{-0.49}$
&&-15&$^{+42}_{-21}$&$^{+67}_{-40}$&$^{+84}_{-55}$
&&0.28&\multicolumn{3}{c}{(fixed)}\\
\\
\textit{$V\,\sin\,I$ Prior on, 2.2 km\,s$^{-1}$} \\
&1.37&$^{+0.35}_{-0.17}$&$^{+0.66}_{-0.35}$&$^{+0.95}_{-0.55}$
&&-44&$^{+98}_{-14}$&$^{+106}_{-19}$&$^{+120}_{-30}$
&&0.28&\multicolumn{3}{c}{(fixed)}\\
\\
\hline
\end{tabular}
\end{table*}

In order to lighten the text, here are placed the results of the Markov Chains using various starting hypotheses and from which we estimated upper limits on eccentricity and long term radial velocity trends. 

Table~\ref{tab:comp} compares the results in $\chi^2$ by instrument and during the Rossiter-McLaughlin effect for eights chains. From this we concluded that the eccentric orbit model is compatible with the circular, that a drift that can be fitted is statistically indistinguishable from zero, but that we detect an anomaly in the reflex Doppler motion, corresponding to the location where the Rossiter-McLaughlin effect is expected. 

Table~\ref{tab:degen} explores the parameters issued from chains where the Rossiter-McLaughlin effect was fitted in order to make sense of the results and see their dependency on initial hypotheses.


\begin{table*}
\caption{The HARPS data, comprising the Rossiter-McLaughlin effect.}\label{tab:harps}
\begin{tabular}{llllll}
\hline
\hline
JBD-2\,450\,000 & RV & 1\,$\sigma$ & BS span & FWHM & Exposure \\
(days)&(km\,s$^{-1}$)&(km\,s$^{-1}$)&(km\,s$^{-1}$)&(km\,s$^{-1}$)&(s)\\
\hline

55184.536563&	5.73155	&0.00360	&0.00524	&6.17685	&1800\\
55184.589179&	5.70957	&0.00791	&-0.01091&	6.12089	&600\\
55184.597270&	5.71090	&0.01113	&0.00694	&6.14773	&500\\
55184.603832&	5.71490	&0.00889	&0.03064	&6.14888	&600\\
55184.610626&	5.71436	&0.00718	&0.02237	&6.14842	&600\\
55184.618080&	5.70599	&0.00754	&0.02609	&6.13617	&600\\
55184.625326&	5.71104	&0.00669	&0.02865	&6.16877	&600\\
55184.632583&	5.71453	&0.00664	&-0.00662&	6.14420&	600\\
55184.639956&	5.70352	&0.00658	&0.00463	&6.13687	&600\\
55184.647270&	5.71617	&0.00673	&0.01262	&6.14405	&600\\
55184.654655&	5.71182	&0.00693	&0.01697	&6.13470	&600\\
55184.661704&	5.70641	&0.00730	&0.01063	&6.15581	&600\\
55184.669227&	5.69401	&0.00835	&0.00671	&6.13308	&600\\
55184.676634&	5.69442	&0.00714	&0.01096	&6.16352	&600\\
55184.683533&	5.68357	&0.00752	&-0.00056&	6.12561	&600\\
55184.691357&	5.68068	&0.00853	&0.02552	&6.14308&	600\\
55184.698822&	5.68454	&0.00769	&0.01908	&6.18173	&600\\
55184.705859&	5.67920	&0.00704	&0.01698	&6.13790	&600\\
55184.713186&	5.66648	&0.00745	&0.01872	&6.14939	&600\\
55184.720582&	5.66897	&0.00667	&0.00385	&6.18264	&600\\
55184.727897&	5.67299	&0.00695	&-0.01616&	6.11436	&600\\
55184.735142&	5.67307	&0.00746	&0.01983	&6.15680	&600\\
55184.742538&	5.68932	&0.00762	&0.02362	&6.10253	&600\\
55184.749853&	5.67476	&0.00822	&0.00608	&6.11238	&600\\
55184.757041&	5.67851	&0.00813	&0.03817	&6.15604	&600\\
55184.764495&	5.67720	&0.00755	&0.00056	&6.13815	&600\\
55184.796046&	5.65836	&0.00524	&0.00142	&6.15492	&900\\
55184.835306&	5.64075	&0.00274	&0.00892	&6.15358	&1800\\
55219.536168&	5.81445	&0.00342	&0.00899	&6.14389	&1200\\
55220.604410&	5.55612	&0.00401	&0.01814	&6.13610	&1200\\
55223.736286&	5.54403	&0.00511	&0.01457	&6.15457	&1200\\
55224.642883&	5.75107	&0.00457	&-0.01011&	6.14606	&1200\\
55233.567511&	5.78692	&0.00566	&0.02799	&6.15914	&1200\\
55234.588418&	5.73021	&0.00741	&0.01902	&6.17153	&1200\\
55236.562452&	5.79533	&0.00505	&-0.00451&	6.14023&	1200\\

\hline
\end{tabular}
\end{table*}

\begin{table*}
\caption{The CORALIE data.}\label{tab:Coralie}
\begin{tabular}{llllll}
\hline
\hline
JBD-2\,450\,000 & RV & 1\,$\sigma$ & BS span & FWHM & Exposure \\
(days)&(km\,s$^{-1}$)&(km\,s$^{-1}$)&(km\,s$^{-1}$)&(km\,s$^{-1}$)&(s)\\
\hline

54709.856049	&5.79531	&0.03672	&-0.03025&	7.78551	&1801\\
54721.862238	&5.83315	&0.03138	 &0.02394	&7.72740	&1801\\
54725.881413	&5.61634	&0.02144	 &0.09526	&7.72961	&1801\\
54726.883360	&5.72078	&0.01895	 &0.04365	&7.79082	&1801\\
54729.893416	&5.72868	&0.01955	&-0.00197&	7.68800	&1801\\
54772.702590	&5.64515	&0.01884	 &0.04254	&7.54286	&1801\\
54826.636491	&5.59334	&0.01734	&-0.05022&	7.66764	&1801\\
54839.638483	&5.78332	&0.01936	 &0.01223	&7.56820	&1801\\
54840.624884	&5.50675	&0.02066	 &0.01755	&7.69087	&1801\\
54853.659660	&5.74287	&0.01743	&-0.03411&	7.58616	&1801\\
54854.753643	&5.71425	&0.01834	 &0.00393	&7.64011	&1801\\
54855.634491	&5.52361	&0.01462	 &0.00976	&7.58558	&1801\\
54856.788275	&5.80608	&0.01708	 &0.01947	&7.66122	&1801\\
54858.748831	&5.54358	&0.02287	&-0.00449&	7.69812	&1801\\
54859.565580	&5.76518	&0.01689	&-0.02529&	7.58044	&1607\\
54860.570382	&5.76323	&0.01486	&-0.06345&	7.63308	&1801\\
54861.608724	&5.53730	&0.01733	 &0.03321	&7.71663	&1801\\
54862.597946	&5.78893	&0.01895	&-0.00452&	7.63392	&1800\\
54865.615004	&5.79862	&0.01535	&-0.06106&	7.65408	&1801\\
54879.698626	&5.59161	&0.02197	 &0.00949	&7.63710	&1801\\
54880.740601	&5.79544	&0.02157	 &0.09068	&7.65103	&1801\\
54881.724950	&5.56329	&0.01882	 &0.00371	&7.60894	&1801\\
54882.713789	&5.61188	&0.01960	 &0.02774	&7.65983	&1801\\
54884.688201	&5.59052	&0.02050	&-0.02699&	7.65376	&1801\\
54885.588094	&5.58612	&0.01961	 &0.02652	&7.63523	&1801\\
54886.669475	&5.79792	&0.01817	&-0.04049&	7.73384	&1800\\
55068.917278	&5.82103	&0.01781	&-0.01827&	7.55927	&1801\\
55096.888493	&5.54043	&0.01820	 &70.04175	&7.58449	&1801\\
55100.879182	&5.73334	&0.02419	 &0.02814	&7.57866	&1801\\
55126.839192	&5.54565	&0.01903	&-0.05241&	7.63741	&1801\\
55186.714652	&5.79927	&0.01229	 &0.03494	&7.59466	&1801\\
55190.804312	&5.60154	&0.01458	 &0.03131	&7.65782	&1801\\
55192.795978	&5.82484	&0.01197	&-0.00958&	7.62474	&1801\\
55193.737931	&5.59628	&0.01294	 &0.05715	&7.59687	&1801\\
55194.799990	&5.60405	&0.01355	 &0.04516	&7.60994	&1801\\
55195.828656	&5.80461	&0.01274	 &0.01753	&7.62373	&1801\\
55291.610872	&5.53752	&0.01454	&-0.00361&	7.68037	&1801\\
55294.539316	&5.51628	&0.01558	&-0.00891&	7.63736	&1801\\
\hline
\end{tabular}
\end{table*}

\end{document}